# Surface plasmon enhanced fast electron emission from metallised fibre optic nanotips

Sam Keramati[1*], Ali Passian[2,3*], Vineet Khullar[2], Joshua Beck[1], Cornelis Uiterwaal[1] and Herman Batelaan[1*]

[1]Department of Physics and Astronomy, University of Nebraska-Lincoln, Lincoln, Nebraska 68588, USA
[2]Quantum Information Science Group, Oak Ridge National Laboratory, Oak Ridge, Tennessee 37831, USA
[3]Department of Physics and Astronomy, University of Tennessee, Knoxville, Tennessee 37996, USA

*E-mail: sam.keramati@huskers.unl.edu (S K); passianan@ornl.gov (A P); hbatelaan@unl.edu (H B)

**Abstract**

Physical mechanisms of electron emission from fibre optic nanotips, namely, tunnelling, multi-photon, and thermionic emission, either prevent fast switching or require intense laser fields. Time-resolved electron emission from nano-sized sources finds applications ranging from material characterisation to fundamental studies of quantum coherence. We present a nano-sized electron source capable of fast-switching (≤1 ns) that can be driven with low-power femtosecond lasers. The physical mechanism that can explain emission at low laser power is surface plasmon enhanced above-threshold photoemission. An electron emission peak is observed and provides support for resonant plasmonic excitation. The electron source is a metal-coated optical fibre tapered into a nano-sized tip. The fibre is flexible and back illuminated facilitating ease of positioning. The source operates with a few nJ per laser pulse, making this a versatile emitter that enables nanometrology, multisource electron-lithography and scanning probe microscopy.

Keywords: fibre optic nanotip, surface plasmon resonance, multi-photon emission, above-threshold emission

## 1. Introduction

Pulsed laser-induced electron emission from metallic nanotips is emerging as a platform for nanometrology, with applications in time-resolved microscopy, electromagnetic field sensing, and fundamental studies of matter-wave interference and quantum optics [1—6]. An important advantage of such nanoscale point-like emitters over the traditional photocathodes is their large spatial coherence [7—9]. Among the numerous applications of sources of free electrons that operate in the cold-field emission regime, the scanning probe microscopy (SPM) techniques of transmission electron microscopy (TEM), scanning electron microscopy (SEM), and scanning tunnelling microscopy (STM) are indispensable for material characterisation and nanotechnology. The resolution of complex electron sources has reached the domain that even time-resolved imaging of atomic and molecular processes is studied now [5, 6, 10]. New electron generation mechanisms may enable SPM to reach higher spatial, temporal, and spectral resolution. As a result, the fields of nanometrology and ultramicroscopy seek to achieve reliable fast and ultrafast laser-induced emission from nanotip emitters [11—14]. Unlike externally stimulated nanotips, optical fibre-based nanotips promise a remarkable ease of use by virtue of the exciting fields being confined within the tip (Figure 1(A)). No alignment of a laser focus on the nanotip, and no laser beam viewports into a vacuum system are required, while the fibre having translational degrees of freedom, provides flexibility for the positioning and scanning of the source with sub-nanometre resolution. The problem is that laser light in the tip (Figure 1(B)) needs to be intense to allow for multi-photon emission in the visible regime and can heat and destroy the nanotip metallic coating. UV light induced electron emission may rely on the (single-photon) photoelectric effect, but does not propagate well through narrow fibres. Additionally, chromatic dispersion in optical fibres increases the laser pulse length. The loss of intensity can be compensated for by increasing the laser power, but at the expense of increasing the risk of damage. Furthermore, laser-induced heating followed by thermionic emission is slow. Therefore, emission mechanisms (Figure 1(C)) that lower the required laser power are called for.

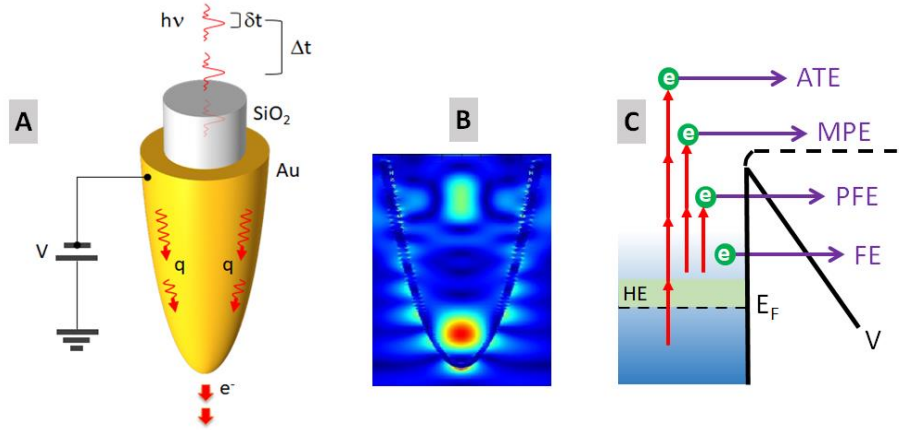

**Figure 1. Electron emission from tapered gold-coated fibre-optic nanotips.** (**A**) Light coupled into a tapered fibre leads to the emission of electrons from the nanotip of the fibre. (**B**) A calculated field intensity pattern at the nanotip is given for a short pulse showing a hot spot at the tip apex. (**C**) Several mechanisms can lead to electron emission. Indicated are: field emission (FE) with tunnelling through a potential barrier (V), photo-field emission (PFE), multi-photon emission (MPE), and above threshold emission (ATE) with a hot electron distribution (HE) above the Fermi energy $E_F$. The Fermi-Dirac distribution is represented with fading above $E_F$. Thermionic emission follows from over the barrier (solid black line) emission from this distribution.

When an electron or photon interacts with metals, collective electronic effects may lead to excitation of surface plasmons with quanta $\hbar\omega_{SP}$, where $\omega_{SP}$ is the frequency of the surface charge density oscillations. Surface plasmons may lower the required laser power for electron emission. For bounded electronic systems, R. Ritchie introduced the concept of surface plasmons to describe the observation of electron energy losses sustained by electrons penetrating metal foils [15]. The field enhancement and confinement associated with plasmon excitation have been used in recent electron photoemission experiments. Examples include surface plasmon resonance (SPR)-enhanced electron photoemission from the corners of Ag nanocubes [16], and the surface of a high-brightness nano-patterned Cu photocathode [17]. In both cases, ultrafast laser pulses, in the long-wavelength regime were used and multi-photon photoemission was identified as the electron emission mechanism. Thermionic emission through IR-absorption was ruled out as a viable channel for the observed enhanced electron emission rates at photon energies a few-fold smaller than the work function of the metals used. In addition, multi-photon and above-threshold photoemissions from the localised plasmonic hot spots of gold bowtie nanostructures were spatially and energetically resolved [18]. SPR-assisted single-photon photoemission in the short-wavelength (UV) regime from a thin film of Al has also been reported [19].

Apart from the need for enhanced emission rates in nanotip emitters some applications require the proximity of the source to a specimen under study. Ultrafast electron point projection microscopy (ePPM) is an important example of such a source-sample configuration. In ePPM, the probe nanotip is brought into close (micrometre) proximity of a specimen, which is to be illuminated by the emitted electrons. The magnifying power of the microscope is given by $D/d$, the ratio of the tip-to-detector distance $D$ to the tip-to-specimen distance $d$. A shorter $d$ thus yields a higher magnification. Since the smallest focal spot size of the ultrafast laser beams is of the same order of magnitude as the minimally desired values of $d$, it is not feasible to irradiate the nanotip apex directly by the laser beam and avoid unwanted scattering and damage to sensitive samples. To alleviate such restrictions, grating coupled gold nanotip emitters have been explored for ultrafast ePPM applications [20—23]. Other important capabilities of the envisioned nano-probes include raster scanning a sample in the near-contact mode of operation for ultrafast patterning and STM purposes, which is not feasible or is at best technically sophisticated in practice due to complications in consistently maintaining the optimal optical focusing of the laser beam on the scanning nano-probe.

Here, we investigate a new type of electron source in which the emitted electron current, emission time, and pulse duration can be controlled optically. The proposed system is based upon a dielectric probe nanotip made by tapering a multimode graded-index (GRIN) optical fibre coated with a thin film of a plasmon-supporting material, such as a noble metal. By coupling a laser beam of suitable wavelength into the fibre, plasmons are excited in the curved metallic film at the tapered end. Theoretical studies [24, 25] of surface plasmons excited at the metal-dielectric interface suggest the corresponding modes undergo a curvature-induced dispersion modification. As a result, useful resonant modes can be extended to the visible wavelength range in order to attain field enhancement at the apex of the probe. Previous fibre-based electron emission studies are limited to the pioneering work by the Miller group [26, 27]. One study invoked a tungsten-coated near-field scanning optical microscope (NSOM) fibre probe [26, 28] to which a 2 mW continuous wave (CW) laser beam from a diode emitting at 405 nm was coupled [26]. The second study involved the gold-coated flat end of a wide-area multi-mode (MM) fibre core irradiated by a femtosecond laser beam at the wavelength of 257 nm [27]. To achieve electron emission in the first study, external DC electric potentials larger than +1 kV were applied between the front face of a microchannel plate (MCP) detector and the fibre tip. Based on the long switching (rise) times on the order of tens of milliseconds to a few seconds, thermal emission was suggested as the most likely mechanism [26]. In the second study, at 257 nm, the photon energy being larger than the gold work function of $\phi \sim 5.2$ eV [29], the emission mechanism was deemed to be the fast process of single-photon photoemission [27]. In the previous work on optical fibres [26, 27], the fibres were not tapered, the end of the fibre did not reach the nanoscale regime, and plasmon resonance was not used as an emission mechanism.

In our work, the proposed SPR-enhanced electron emission capitalises on the availability of resonant modes sustained by the thin nanoscale curved film. The plasmonic mechanism is corroborated by characterising the electron emission and photonic measurements along with theoretical considerations including a finite difference time domain (FDTD) computational analysis. The aim of our article is primarily two-fold: first, we provide evidence that the observed photoemission at low laser powers is assisted by surface plasmon excitation, and second, we demonstrate that fast sub-nanosecond switching is indeed achievable from the proposed electron gun.

## 2. Design of the nanotip emitter

To answer the question whether or not electron emission, and in particular, SPR-assisted electron emission can be observed from a nanoscale fibre tip, we guide our fabrication by the following considerations. First, we note that for photon excitation of plasmons to occur in the tip region, both the energy and momentum of plasmons and photons must overlap. Since photons do not couple to plasmons directly for interfaces of low curvature and without an assisting coupling material, a specific dielectric and metallic material and a nano-scale radius of curvature of the nanotip are chosen. Second, the thickness and morphology of the metallic domain is known to affect the damping and scattering of plasmons. Therefore, our further objective is to obtain reasonable estimates for the electric field given a coating thickness. Surface plasmon dispersion relations provide a means to explore both the material and geometric parameters. The following consideration are further justified with FDTD-computations.

Describing the material with a local complex frequency-dependent dielectric function $\varepsilon(\omega)$, the dispersion relations are calculated using the solutions of the Laplace equation in the quasistatic regime [24]. Modeling both the nanotip and its coating as confocal one-sheeted hyperboloids of revolution, we obtain the non-retarded plasmon dispersion relations [24]. For a multiply $N$-coated single nanotip in vacuum, the quasistatic surface modes of wavevector $q$ are thus described by the frequencies $\omega_i^m(q,\theta)$, where $m = 0, 1, \ldots$ is the azimuthal mode number, $i = 1, 2, \ldots, N$ denotes the number of interfaces, and $\theta = \{\theta_1, \theta_2, \ldots, \theta_N\}$ represents the opening (polar) angles of the various layers "$i$", measured from the symmetry axis of the nanotip. The number of interfaces $N$ thus signifies the number of dispersion relation branches for each $m$, where the boundary conditions are met when the following relation is satisfied

$$\sum_{k=1}^{N} c_k^m(q) \varepsilon^k(\omega,q) = 0. \qquad (1)$$

For a simple solid tip with an opening angle $\theta_1$ and dielectric function $\varepsilon_t$, a single ($N = 1$) branch $\omega^m(q,\theta_1)$ is obtained such that when $\theta_1 \to \pi/2$, one obtains the well-known [24] reduced frequency $\omega^m(q,\pi/2) = \omega_p/\sqrt{2}$, for all $m$, corresponding to the surface plasmons on a cartesian metal half-space. An additional consistency check is that the non-retarded surface plasmon resonance frequency of $\omega_p/\sqrt{2}$ is also reached for large wavevectors $q$. The higher the curvature of the tip $(\theta_t \to 0)$, the more negative the value of the dielectric function of the tip $\varepsilon_t$, resulting in a redshift of the plasmon energies $\omega^m(q,\theta_t) < \omega_p/\sqrt{2}$. For the specific case of $N = 2$, that is, the singly coated tip considered here, the resonance frequencies of the formed thin curved film with dielectric function $\varepsilon_c = \varepsilon(\omega)$ can be shown to be given by the solutions to

$$c_0^m(q) + c_1^m(q)\varepsilon + c_2^m(q)\varepsilon^2 = 0. \tag{2}$$

In the cartesian thin film limit, that is, $\theta_1 \approx \theta_2 \to \pi/2$, we obtain the correct limits $c_0^m(q) \to 1$ and $c_2^m(q) \to 1$, while $c_1^m(q) \to 2\coth q(\theta_c - \theta_t)$. To illustrate the meaning of $q$ for an actual planar thin film model (with thickness $a$), we find that $c_0 = c_2 = 1$ while $c_1(k) = 2\coth ka$, which would correspond to the well-characterised symmetric and antisymmetric charge density oscillations (with momenta $k$) of a vacuum-bounded planar thin film [24]. The choice of the fibre core affects the plasmon damping. A hollow fibre gives no damping. The plasmon frequency redshifts on the thin curved film (with probe tip $\theta_t$ and coating $\theta_c$ angles) with increasing value of the core material dielectric function $\varepsilon_t$ so that an appropriate choice of material has to be made. We found that silica is a good choice to obtain a plasmon resonance in the visible range. From the above equation, the resonance conditions of the tip in the visible spectral range can be simulated. For example, using the dielectric properties of gold and silica [30], several modes can be observed in the visible spectrum for practical angular values (20—60 degrees) and coating thickness (~ 50 nm). Guided by these estimates, we fabricated a series of gold-coated pulled optical fibre nanotips. See the Materials and Methods (Appendix A) for more explanation.

### 3. Experimental setup

The schematic experimental setup used to test our designed fibre nanotip is shown in Figure 2. The fibre tip is mounted inside a high-vacuum chamber using a fibre chuck. A fine piece of Cu wire is used to apply a negative electric bias of $V_{tip} = -50\,\text{V}$ as an accelerating potential to the shank of the metallised fibre tip using an electrical feedthrough on the host vacuum flange. An electron detector at a fixed distance of 1 cm from the apex of the nanotip is used to detect the emitted electrons. A home-built fibre optic vacuum feedthrough guides the laser light into the vacuum chamber whereby end-fire-coupling the light into the 30 cm long fibre tip is accomplished. The fibre part of the nanotip can be made shorter to reduce laser pulse broadening due to optical dispersion. The feedthrough is made using the same type of fibre in order to minimise the scattering losses due to cross-sectional area mismatch of the end-fire-coupling stage.

To measure the laser power input to the nanotip fibre, an auxiliary fibre was used. The power was measured at the terminated end of the auxiliary fibre. A connectorised collimating lens with antireflection (AR) coating spanning the visible range (i.e., fully encompassing the wavelengths employed in this work) was used for power measurements. The average laser power was controlled by a continuously variable neutral density (ND) filter. Based on the fibre transmission specification, approximately 80% of the electromagnetic power measured after the auxiliary fibre reaches the beginning of the taper of the fibre tip. Light leakage from the taper prevents accurate measurement of the fraction of the light that reaches the tip. The light transmission from the fibre tip is approximately 0.3 % of the input as measured with a photodiode.

A coincidence technique was used to explore the switching speed of the electron pulses. The laser pulse triggers a start signal for the measurement of a time window, which is stopped by the detection of the electron pulse. A solid state pumped Ti:Sapphire oscillator followed by a regenerative amplifier and an optical parametric amplifier (OPA) generated the pulsed beam with a pulse duration of 50 fs and a repetition rate of 1 kHz. The wavelength-tunable output beam of the OPA was used to search for the presence of a plasmonic resonance in the electron emission spectrum.

### 4. Results

#### 4.1 Intensity- and wavelength-dependent electron emission

To study the dependence of the electron emission on the optical field intensity, we employed a ND-wheel to control the intensity of the input laser beam. This ND-wheel is rotated with a constant rotation speed and its direction of rotation is reversed periodically. The result is shown in Figure 3(A), where the electron emission count rate $C$ is monitored at the two well-separated wavelengths of $\lambda_{ML} = 500$ nm and 660 nm. Although the $x$-axis is displayed linearly in time, the ND-wheel transmission depends exponentially on its angle. To investigate whether the emission exhibits a power law, the vertical axis is displayed on a logarithmic scale in Figure 3(B). The electron emission is governed by a power law, $C \propto I^{\bar{n}}$, where $\bar{n}$ is the power-law order.

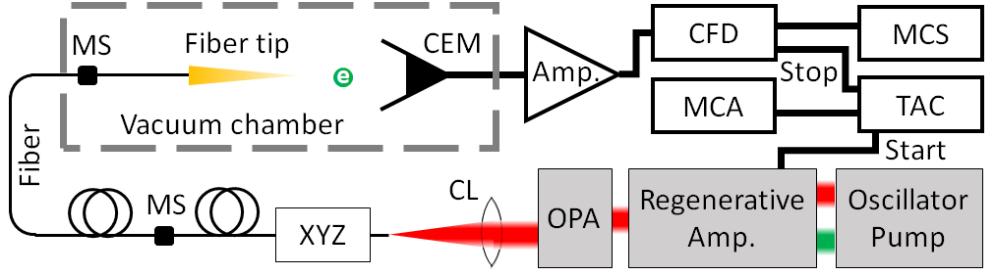

**Figure 2. Experimental schematic.** Light from a femtosecond laser oscillator and a pump laser drive a regenerative amplifier. The wavelength-tunable light from an optical parametric amplifier (OPA) is focused onto a fibre with a coupling lens (CL). The fibre is positioned with a translational XYZ stage. The mating sleeve (MS) between the first and auxiliary fibres allows for power measurements. The auxiliary fibre passes into the vacuum system, where a second mating sleeve facilitates switching between different nanotip fibres. Electrons emitted from the nanotip fibre are collected with a channel electron multiplier and electronically processed (see text).

Having characterised the emission at the special (on- and off-resonance) wavelengths above, we now seek to obtain the broader spectral properties of the emission power-law order. The result is shown in Figure 3(C), where $\bar{n}$ is displayed as a function of the OPA wavelength. A wavelength-dependence is observed with a peak at ~ 660 nm. The red solid curve in the plot represents a model based on resonance-enhanced above-threshold emission (ATE), which will be discussed in the theoretical model section below. To further explore the emission parameter dependencies, we visualise the emission rate as a function of both OPA wavelength and average power, as shown in the contour plot of Figure 3(D). The contours represent the emission count rate as obtained from the intensity data at all 13 wavelengths by extrapolating their fit functions. The count rate maximises at the resonance wavelength. For an average power of 12 μW, indicated by the dashed line in the contour plot, a spectral peak for the measured electron count rate is found (Figure 3(E)), which corresponds to the peak of $\bar{n}$ in Figure 3(C). The green solid curve in Figure 3E is the result of the same ATE model with the same parameters as used for the result in Figure 3(C). The number of wavelengths used is limited due to the need to realign the OPA for each wavelength setting. The data presented in Figure 3(C & E) is the average of three data runs, two with 13 and one with 4 wavelengths measurements.

*4.2 Model of the emission process*

The observed power law suggests that part of the emission mechanism relies on multi-photon electron emission. The emission count rate can be described by

$$C = N \sum_n a_n I^n p_n \equiv N \sum_n C_n, \qquad (3)$$

where $N$ is an overall scaling factor, $a_n$ is a transition probability, $p_n$ is a population, and $I$ is the intensity. This equation motivates Figure 1(C). For example, considering the four-photon ATE process, each red arrow adds a factor $I$ to the count rate. The process starts from states in the Fermi-Dirac distribution with population $p_n$, while the total process has some quantum mechanical transition probability indicated with $a_n$. The transition probability $a_n$ has the property $a_{n+1}/a_n \ll 1$ indicating that higher-order processes are less likely, as appropriate for a perturbative series for which $a_n \propto \xi^n$, where $\xi$ is a constant [31]. The intensity is given by

$$I = I_0 \left[ 1 + \delta \exp\left(-(\lambda - \lambda_0)^2 / \lambda_w^2\right) \right], \qquad (4)$$

where the peak laser pulse intensity, $I_0 = P/(\Delta t \pi r^2 f_{rep})$, depends on the average power, $P$, the laser pulse duration, $\Delta t$, the radius of curvature of the nanotip, $r$ (~ 50 nm), and the laser repetition rate, $f_{rep}$ (= 1 kHz). The SPR wavelength is $\lambda_0 = 660$ nm and the resonance width is $\lambda_w = 30$ nm. A modest field enhancement factor of $\delta = 3$ provides reasonable agreement with the experimental data. Both the order and the count rate at the resonance restrict the magnitude of the enhancement factor. The laser linewidth (Figure 4(A)) is three times narrower than the resonance width and not taken into account. The population $p_n$ is that of the state from which the multi-photon excitation starts. A value approximately given by the Fermi-Dirac distribution would be expected for a metal,

$$p_n = \frac{1}{\exp\left(\frac{\phi - n\hbar\omega}{k_B T}\right) + 1}, \tag{5}$$

where the workfunction $\phi = 5.25$ eV for gold is lowered by the energy of the number $n$ of photons needed for the multi-photon excitation process. The population equals one for states below the Fermi energy (Figure 1(C)), while the energy of three photons of wavelengths ranging from 500 to 700 nm is needed to get multi-photon emission (MPE) from these states. A surprising observation is thus, that the order parameter hovers around two for wavelengths from 500 to 620 nm (Figure 3(C)), indicating that a two-photon process dominates. Moreover, the emission rate does not drop rapidly by orders of magnitude as the wavelength increases in this same range (Figure 3(E)). The Fermi-Dirac distribution would predict such a behaviour above the Fermi level. We assume that the first transition in the three-photon process saturates, which would make the population independent of wavelength and the order value equal to two [32]. We model this by replacing $n \rightarrow n+1$ in Eq. 5. Distributions of hot electrons with a width of 1-2 eV have been predicted for a gold surface by laser excitation providing a candidate for the saturated state [33].

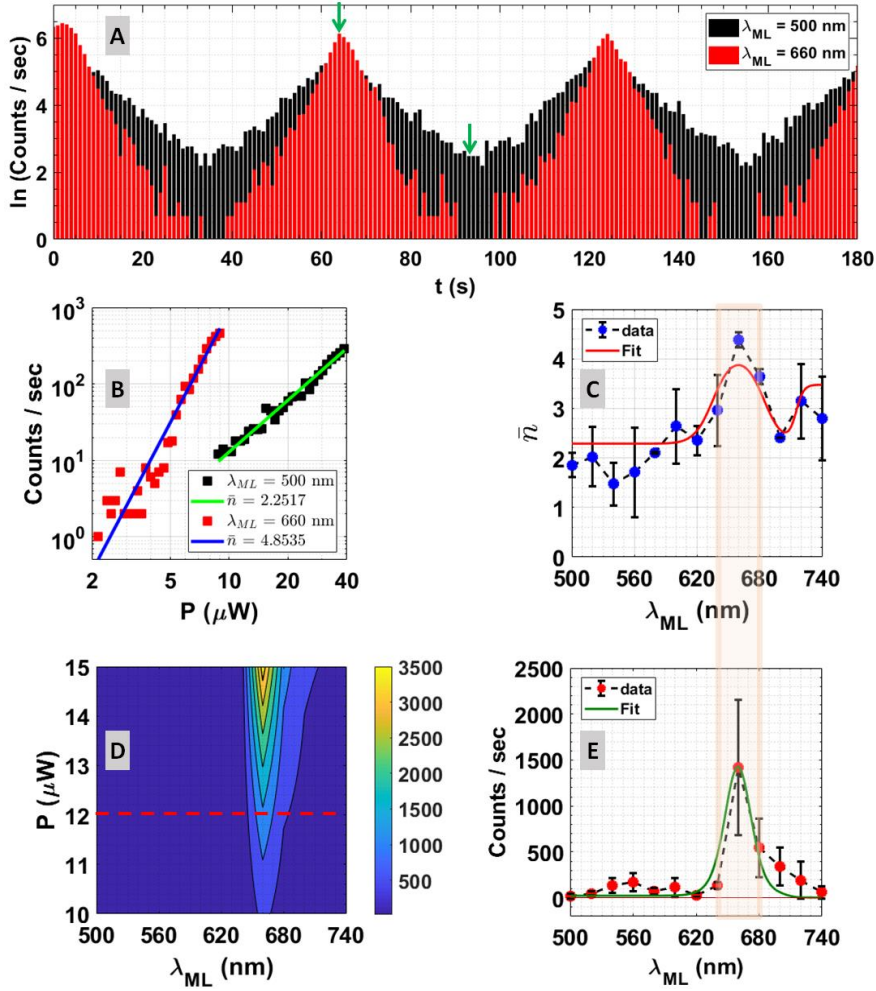

**Figure 3. Pulsed laser photoemission wavelength dependence.** (**A**) Two examples of electron emission as a function of laser intensity (log-log scale) are given at 500 and 660 nm. The laser intensity is ramped up and down several times. (**B**) The two examples in the intensity interval between the green arrows in panel (A) are fitted to a power law, $C \propto I^{\bar{n}}$, with power-law order $\bar{n}$. The fit is a straight line in the log-log representation with slope $\bar{n}$. (**C**) The power-law order is shown as a function of wavelength. Errorbars indicate one standard deviation. A wavelength-dependent resonance peak at ~ 660 nm is manifest. The red solid curve is a model based on resonance-enhanced above-threshold emission (ATE). (**D**) The electron emission count rate at the thirteen wavelengths are displayed in a contour plot using the extrapolated fit functions. The count rates maximise at the resonance wavelength. (**E**) The electron emission count rate (red dots) shows a peak (along the dashed line in panel (D)) at the same wavelength as in panel (C). The green solid curve is produced with the ATE model and the same parameters as used in panel (C). The dashed line guides the eye in panels (C) and (E).

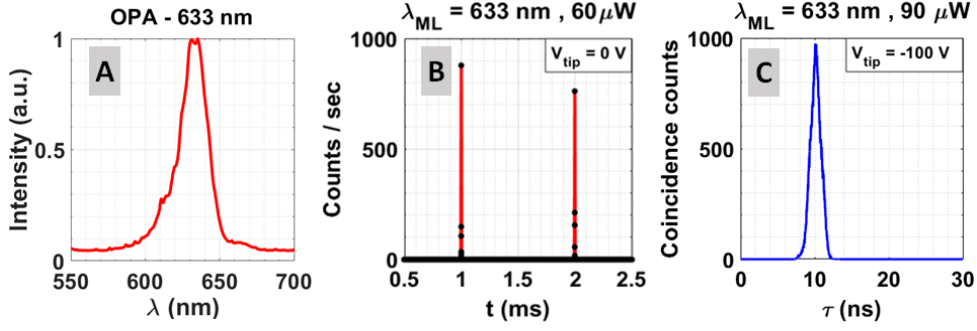

**Figure 4. Optical switching of electron emission by low-power femtosecond lasers. (A)** The normalised output spectrum of a femtosecond pulse from an optical parametric amplifier (OPA) with the central wavelength set at 633 nm is shown. **(B)** The pulsed electron emission follows the laser repetition time interval of 1 ms. The average power is 60 μW, and $V_{tip}$ can be set at 0 V (B) or the electrons can be accelerated to 100 eV (C). **(C)** The coincidence time spectrum between the laser pulse and the subsequent electron detection events is shown as a function of the time delay between the two. The peak width is limited by the time resolution of the detector. This sets an upper limit of ~ 1 ns for the emission time duration.

The order parameter is obtained from Eq. 3 by determining the slope as a function of intensity on a log-log scale

$$\bar{n} = \frac{I}{C}\frac{dC}{dI} = \frac{I}{C}\sum_n n a_n p_n I^{n-1} = \frac{1}{C}\sum_n n C_n \qquad (6)$$

Upon inspection of the right-hand side, this equation yields the average order $\bar{n}$. It has no further adjustable parameters and the model is strongly constrained by the combination of the experimentally measured order parameter and emission rate (even when the experimental data is sparse). For the model curves (Figure 3), the parameters are $a_1 = 1 W^{-1} m^2 s^{-1}$, $a_2 = 1.7 \times 10^{-16} W^{-2} m^4 s^{-1}$, $a_3 = 3.0 \times 10^{-32} W^{-3} m^6 s^{-1}$, $a_4 = 1.0 \times 10^{-47} W^{-4} m^8 s^{-1}$, $a_5 = 7.0 \times 10^{-63} W^{-5} m^{10} s^{-1}$, and $N = 1.3 \times 10^{-13}$.

### *4.3 Alternative emission mechanisms*

The objective of this study is to find a flexible electron source that lowers the amount of laser power needed, so that the source is less likely to suffer damage. The mechanism discussed above satisfies this objective. It is, however, important to look at alternative mechanisms that could explain the resonance without providing protection from damage. In particular, no gain has been made if a plasmonic resonance or an intermediate state resonance heats the nanotip and drives thermionic electron emission or photo-assisted thermionic emission. The strong non-linearity of thermionic emission could in principle change the order parameter as a function of wavelength in the presence of a resonance. The question then is if our data rules out thermionic emission. The intensity-dependent data requires an assumption on how the tip temperature depends on laser power. Assuming the temperature is linearly dependent on the laser power delivered to the tip, that is $T(P) \propto P$, the Richardson-Laue-Dushman (RLD) equation [29],

$$C_{RLD} = \left(\frac{\pi r^2}{q_e}\right) \cdot A_{RLD} T(P)^2 \exp\left(\frac{-\phi}{k_B T(P)}\right), \qquad (7)$$

predicts an electron emission that is much more strongly dependent on power than a power law with order two. This would be inconsistent with our data for wavelengths in the 500-620 nm range. Moreover, a resonance would change the order $\bar{n}$ and the emission much more than observed, as the laser power now enters not only in the exponent of the power, but also in the exponential. A model similar to the photon-enhanced thermionic emission (PETE) mechanism [34] would modify Eq. 7 by multiplication with $I^n$ and would lower the workfunction as in Eq. 5. These models all lead to variation in emission and order that are too large. Reducing the resonance to a very weak one can give some reasonable agreement with the electron emission but at the expense that the order parameter loses its resonance. The simultaneous presence of a resonance in the order and emission rate strongly constrains the possible emission mechanism.

*4.4 Pulsed electron emission*

The switching speed of the electron source is relevant for applications in time-resolved studies. Here, a modest time resolution is investigated with an eye on ruling out the alternative mechanism of thermionic emission assisted by plasmon-enhanced heating of the Au layer. A previous study on gold nanowire damage predicts a cooling time of ~ 10 µs [35]. The silica fibre core in our gold-coated fibre tip is a thermal insulator, which hampers cooling as compared to solid metal nanotips, consistent with studies of plasmon decay in the Kretschmann configuration [36] and nanoparticles [37]. To study the electron pulse duration a femtosecond laser pulse was launched into the optical fibre.

The modal dispersion in GRIN fibres is minimal and laser pulse broadening is dominantly caused by chromatic dispersion [27]. An estimate using the dispersion relation for the fibres from which we fabricated our probes yields 160 fs nm$^{-1}$ m$^{-1}$ for the chromatic dispersion at a wavelength of 750 nm. For the 3 m long fibre in our experiment involving the auxiliary and feedthrough fibres and the fibre tip itself, we estimate a typical ~ 5 ps broadening in the visible spectrum for the OPA pulses reaching the metallised nanotip. The resulting pulse stretch by about two orders of magnitude leads to a reduction of the peak optical electric field at the emitting tip. The electron photoemission is measured as a function of time for an average power of 60 µW (Figure 4(B)). The power is higher than that used for the data in Figure 3(E), which might have increased a thermionic contribution to the signal. Coincidence timing between the laser pulse and electron pulse was used [38] to obtain a temporal peak width of ~ 1 ns (Figure 4(C)). The observed width is attributed to the electron detector temporal resolution and sets the experimental upper limit of the electron emission pulse duration. Thermionic emission is unlikely as no cooling is observed. For the ATE model, the emission mechanism is expected to reach into the femtosecond domain. It should thus be possible to reach this domain for a 1 cm long fibre.

*4.5 Photonic characterisation*

The fabricated nanotip emitters were designed to allow optical excitation of surface plasmons near the tip apex (radius of curvature ~ 50 nm, and metallic thickness ~ 50-100 nm). As schematically depicted in Figure 1, the driving laser light, that reaches the metallised tip through the fibre core, excites surface plasmons with wavenumber $q$. As the plasmon dispersion relations predict that the curvature of the Au coating at the apex furnishes available momentum for the plasmon-photon coupling, it also provides a decay channel. Therefore, the radiative decay of plasmons is expected near the tip apex, including any resonances that may be present in a given spectral window. To determine the photonic response of the fabricated nanotips, we employed a light source with a spectrum in the visible range, a monochromator (Cornerstone, Newport), and a photomultiplier tube (PMT) detector (Hamamatsu). Both a mercury arc lamp (Newport) and a supercontinuum white light laser (Leukos) source were tested for performance comparison. The output of the monochromator was coupled into the fibre and the nanotip was inserted into the PMT housing. A mechanical chopper and a lock-in amplifier were used to measure the emitted light from the nanotip.

The radiative plasmon decay at the nanotip apex detected (in the far-field) by the PMT is expected to be consistent with the electron emission spectral response of the fibre tip. Thus, at wavelengths where electrons are emitted resonantly from the nanotip, plasmon excitation and decay are expected also to exhibit a resonance behaviour for an identical fibre nanotip. Considering the many parameter dependencies, modest agreement is expected between experiment and the model. The resonant wavelength is sensitively dependent on surface curvature, roughness, and coating thicknesses and materials. Therefore, variations in the outcome of our current fabrication parameters can lead to tens of nanometre variation in the resonance position and width [24]. The fabrication process is sufficient to meet our objective of shifting the resonance peak into the visible domain. Both the theoretical analysis and the optical experiment show a peak in the visible range (Figure 5(A)). An earlier computer simulation of the optical field enhancement that was also attributed to the plasmon excitation at a similar Au nanostructure (with a radius of curvature of 10 nm), reported a theoretical peak consistent with our observed resonant wavelength [39].

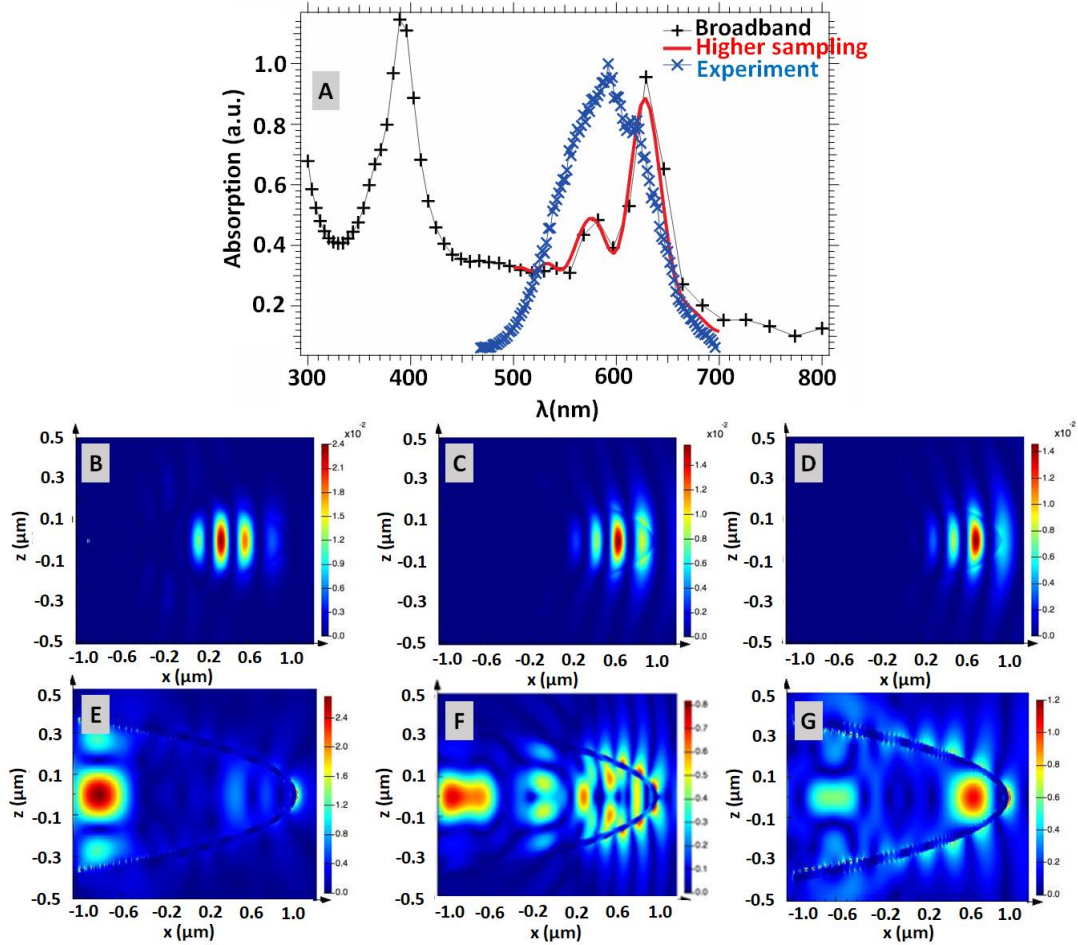

**Figure 5. FDTD simulation of the plasmonic response of the nanotip.** (**A**) Computationally determined absorption spectrum of the nanotip and comparison with the experimentally observed spectral peak. Absorption spectra have some variation in a range of similarly manufactured nanotips. Peak positions can shift by tens of nanometres due to details of the nanotip morphology. (**B-D**) Computed instantaneous field distribution for a tapered but uncoated fibre tip is visualised to display the dominant excited modes. Light from the 2 fs dipole source is transmitted to vacuum without loss. This is expected and indicates that no energy is available for plasmon excitation. (**E-G**) Instantaneous field distribution for the tapered and gold-coated fibre tip at three successive times. The dipole source is initiated in panel (E), and panels (F) and (G) are at 17 fs and 21 fs, respectively. The hot spot in panel (G) and the phase shift between the waves inside and outside the fibre are indicative of the surface plasmon resonance. Additionally, the initial emission has died out, while an enhanced electric field is present near the apex. This is in stark contrast with the distributions in panels (B-D) for the uncoated tapered fibre.

To model the observed photonic signal, in principle the quasistatic formulation discussed earlier can be used to obtain the near-field distribution corresponding to a given set of eigenvalues ($m,q$); the higher the plasmon momentum $q$, the higher the charge density oscillation along the hyperboloidal surface. However, to avoid further analytical complexities [25], in order to investigate whether the experimentally observed photonic spectral peak can be predicted theoretically on the basis of plasmon excitation near the tip apex, we further employed the FDTD technique to numerically compute the fields. Within the FDTD computational domain, a 2 μm long probe with an apex radius of curvature of 50 nm was created for a silica core and a 50 nm thin film of gold. To obtain the response of the probe, a source field, here a 2 fs pulse, was created from within the $SiO_2$ core and propagated to the Au domain, as visualised in Figure 5(E—G). This can be achieved by incorporating a dipole radiation field near the input of the fibre on the symmetry axis. Any ensuing photon-plasmon coupling that occurs exhibits a resonance structure that is consistent with the normal modes of charge density oscillations on the probe surface, as labelled by the quantum numbers ($m,q$). For comparison, pulse propagation is computed for a tapered but uncoated fibre tip (Figure 5(B—D)). Light is transmitted to vacuum without loss. No energy causes plasmon excitation. Pulse propagation in the Au-coated fibre tip is shown at three successive times (Figures 5(E—G)). The hot spot in panel (G) and the phase shift between the waves inside and outside the fibre are indicative of the surface plasmon excitation.

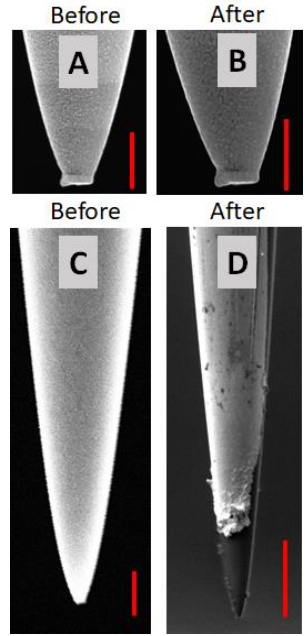

**Figure 6. Scanning electron microscope images of nano-fibretips.** A nanotip with a recognizable structure was selected for a damage test. The SEM image of this tip before **(A)** performing the damage test is shown; scale bar = 500 nm. **(B)** SEM image of the same tip is shown after driving it with the OPA beam at $\lambda_{ML}$ = 500 nm and 14 nJ per laser pulse (at a repetition rate of 1 kHz and 14 µJ average power, that is a bit above the average power used to take the data shown in Figure 4) for 2 minutes giving rise to an average electron count rate of 150 cps at $V_{tip}$ = 0. No damage in the coating is visible; scale bar = 500 nm. **(C, D)** SEM images of the fibre tip showing a before and after image with damage; most of the coating at the fibre tip is removed. The scale bars are 1 µm and 5 µm, respectively. During experiments that span days of operation, the onset of DC field emission tip voltage is checked at regular intervals and remains constant. As the laser intensity is increased in the experiments, at some point in time the photoemission abruptly ceases, and the onset DC field emission potential rises by several hundred volts. Subsequent SEM inspection reveals nanotip coating damage at that point as in panel (D). No emission is obtained from the silica core of the tip without coating.

The computed absorption spectrum, shown in Figure 5(A), exhibits two main features; a 390 nm spectral peak corresponding to a resonant excitation of surface plasmons when the dipole moment of the exciting source is perpendicular to the symmetry axis of the probe, and a 626 nm peak corresponding to plasmon excitation with the field of a dipole that has its moment parallel to the symmetry axis of the probe. The experimentally observed absorption peak is obtained using a gold-coated fibre tip. The tip was similarly prepared as the sharp nanotip whose SEM image is in general agreement with the present simulation result.

## *4.6 Monitoring nanotip damage*

A specially selected fibre tip with a recognizable shape was used for a damage test (Figure 6(A)). The OPA output at $\lambda_{ML}$ = 500 nm and $P$ = 14 µW was used to drive the fibre tip for 2 minutes. A constant detection rate of 150 cps was observed. The high-resolution SEM images of this tip taken before and after this test are shown in Figure 6(A & B), respectively. No damage to the coating is observed. Throughout the experiments, progressively higher intensities and tip voltages towards the end of each set abruptly disrupt the coating. The SEM image of the fibre tip (Figure 6(C)) is shown before laser exposure. The damage is monitored at regular intervals by observing the voltage at which DC-field emission starts. A typical value of approximately −450 V is observed for days, while after a rise in laser power the required DC voltage needed for field emission exceeded −1 kV. Upon inspection of the SEM image damage is found of which Figure 6(D) is a typical example. The fact that the nanotips with a damaged coating do not photoemit or DC emit, indicates that the electron emission originates from the sharp metal coating near the apex and not the heated glass. This is further illustrated by the FDTD simulation results of the instantaneous radiation field propagating in a tapered uncoated fibre shown in Figure 5.

# 5. Discussion

The presented measurements demonstrate the first observation of electron emission from an optically driven nanoscale fibre tip coated with a thin film of gold. The electron emission can be driven with a low-power femtosecond laser oscillator. The proposed photonic fibre-optics approach, as opposed to free-space scattering, is therefore a viable path to electron generation. The described fibre-based source can be readily configured to supply electrons with the same degree of 3D spatial manoeuvrability as in the SPM techniques. That is, fast electron pulses can be delivered from a probe with nanometre lateral and vertical resolution to specific sites of a given specimen. Similar to STM operation, it is expected that the proposed electron nanotip source can operate under ambient conditions, extending its use to a variety of nanometrology applications.

Plasmon bands in other nanostructures of similar size and curvature also yield a spectral peak associated with excitation of specific normal modes of the surface charge density in the gold thin film of a given curvature [24, 25]. From the presented analytical calculations of the nonretarded plasmon dispersion relations for a gold-coated dielectric probe, the availability of several symmetric and anti-symmetric modes for the charge density oscillations can be observed. The computationally determined absorption spectrum of the model exhibits two major resonance peaks associated with plasmon excitation with longitudinally and transversely polarised field components. One of the predicted peaks led to the current study. A search for the ~ 400 nm plasmonic resonance was not attempted in this study as the employed OPA does not support laser emission in this wavelength range, but may yield strong electron emission at even lower laser powers as the power-law order of photo-emission may be lower for more energetic photons. Higher electron emission rates with a further reduced risk of damage could be the result.

Although a quantum treatment [25] of the surface charge density and their interaction with a photon field can be carried out to obtain the radiative decay rate of the plasmons on the surface of the nanotip, our analytical and computational calculations were aimed for our presented proof-of-principle design stage. Further investigation is warranted for a more comprehensive model to account for the photon-plasmon coupling and plasmon-electron interaction.

In conclusion, the demonstrated optically controlled electron emission from a curved gold thin film is of potential for nanometrology, as well as for fundamental studies of photon-electron, electron-electron, electron-plasmon interactions at interfaces. Our preliminary results, supporting a plasmon-enhanced emission mechanism, constitute a proof-of-principle. Further study is useful to establish the detailed role of plasmons and to understand the parameter dependencies, including various material deposition, multilayer deposition, thickness, size and tip curvature, and excitation pulse properties. In light of the progress in quantum materials [40] such as those exhibiting topological behaviour [41], or in 2D materials, e.g., involving graphene plasmonics [42, 43] and plasmonic properties of $MoS_2$ [44], as well as in quantum coherence [45], more advanced and versatile configurations of the presented nanotips may be explored.

## Appendix A. Materials and methods

We tapered our GRIN MM optical fibres (Corning, InfiniCor 600) using a commercial micropipette puller (Sutter Instrument Company). The fibres have a nominal core diameter of 50 μm, a cladding diameter of 125 μm, an outer protective polymer coating of diameter 250 μm, and a numerical aperture of NA = 0.20. An alternative approach to make such nanotips, often for NSOM applications, is chemical etching [46]. The tips were subsequently coated first with 4 nm of Cr and then with 100 nm of Au on top of Cr using electron beam deposition. Cr is used to make the interfaces more robust as Au alone would not adhere well to the glass of the fibre.

The laser beam coupling into the fibre proceeded as follows. The open end of an auxiliary fibre is inserted into a 250-μm ferrule of a fibre chuck mounted on a 3D micro-stage to bring its coupling face into the laser focus after a convergent coupling lens. Subsequently, the laser beam is end-fire-coupled into a fibre optic feedthrough using a standard SMA905 mating sleeve. The feedthrough itself is a one-meter long optical fibre protected by a 900-μm furcation tube. The tube is further protected by insertion into a 2.3-mm stainless steel tube. To make the feedthrough, a 2.5-mm hole was drilled through the centre of a 2.75-inch vacuum flange and threaded with a steel tube hosting the fibre. Both ends of the tube were connectorised. The drilled hole was sealed with vacuum epoxy (TorrSeal). The experiments were performed at a pressure of ~ $2 \times 10^{-7}$ Torr. High-vacuum-rated SMA-SMA mating sleeves were used inside the vacuum chamber.

Fowler-Nordheim [29] curves for DC-field emission were fitted to determine the tip radius of ~ 50 nm, nominally consistent with the SEM image of the tip.

The output voltage signals of the channel electron multiplier (Dr. Sjuts, Model KBL 510) were amplified with a pre-amplifier (ORTEC, Model VT-120). The amplified signal was subsequently fed into a constant fraction discriminator (CFD) (ORTEC, Model 935) in the updating mode. In this mode of operation, an input signal arriving within a set time window (100 ns in our case) after a preceding one does not generate a new output logic pulse. The CFD output signals were registered and counted using a multichannel scaler (MCS) triggered by the regenerative amplifier reference signal. For coincidence timing experiments, the laser amplifier reference signal triggered the start input of a time-to-amplitude converter (TAC) (ORTEC, Model 567) while the CFD output was fed into the stop signal. A multichannel analyser (MCA) recorded the electron emission timing spectrum.

The peak laser pulse intensity, $I_0 = P/(\Delta t \pi r^2 f_{rep})$, depends on the average power, $P$, the laser pulse duration is $\Delta t$, the radius of curvature of the nanotip is $r$ (~ 50 nm), and the laser repetition rate is $f_{rep}$ (= 1 kHz). For a power of 12 μW and a pulse duration of 5 ps, the intensity is $9.2 \times 10^{14}$ W/m$^2$. The associated electric field is ~ 1 V nm$^{-1}$, which is low compared to the experimental start value of optical field emission for a variety of nanotip emitters [3]. This justifies the use of the word "low" and the assumption that the process involves multiphoton emission.

The FDTD analysis solves the Maxwell equations over a Cartesian mesh (voxels) throughout the modelling structure, where each cell is characterised by the properties of the material it occupies. The properties are the frequency dependent dielectric functions of Au and SiO$_2$. An appropriate conformal mesh is created taking into account the relationship between the time increment, incident field wavelength, and nanostructure size. The final fields are then calculated by spatially staggering the field components on the so-called Yee-cell and updating them via the leap-frog time-marching method [39, 47]. Thus, transient fields can be effectively computed when a suitable source field is incorporated. By obtaining the fields everywhere, the absorption may be calculated from an integration of the loss function over the metallic subdomain.

**Acknowledgements**

A. Passian acknowledges support from the laboratory directed research and development fund at Oak Ridge National Laboratory (ORNL). ORNL is managed by UT- Battelle, LLC, for the US DOE under contract DE-AC05- 00OR22725. The fibre optic probes were fabricated at ORNL. S. Keramati and H. Batelaan acknowledge support by a UNL Collaborative Initiative grant, and by the National Science Foundation (NSF) under the award numbers EPS-1430519 and PHY-1912504. The SEM images were taken at the NanoEngineering Research Core Facility (NERCF), which is partially funded by the Nebraska Research Initiative. We thank Pavel Lougovski for help initiating this project. The authors declare no competing interests.